# PHASE-PRIORITY BASED DIRECTORY COHERENCE FOR MULTICORE PROCESSOR

Gongming Li[1] and Hong An[2]

[1] Department of Computer Science and Technology, University of Science and Technology of China,
lgm@mail.ustc.edu.cn
[2] Department of Computer Science and Technology, University of Science and Technology of China,
han@mail.ustc.edu.cn

## *ABSTRACT*

*As the number of cores in a single chip increases, a typical implementation of coherence protocol adds significant hardware and complexity overhead. Besides, the performance of CMP system depends on the data access latency, which is highly affected by coherence protocol and on-chip interconnect. In this paper, we propose PPB (Phase-Priority Based) cache coherence protocol, an optimization of modern directory coherence protocol. We take advantage of the observation that transient states occur in directory coherence protocol, resulting in some unnecessary transient states and stalling. PPB cache coherence protocol decouples a coherence transaction and introduces the idea of "phase" message. This phase is considered as the priority of the message. Additionally, we also add new priority-based arbitrators in on-chip network to support PPB cache coherence protocol. This mechanism in on-chip network can support effective cache access, which makes the on-chip network more efficient. Our analysis on an execution-driven full system simulator using SPLASH-2 benchmark shows that PPB cache coherence outperforms a MESI based directory, and the number of unnecessary transient states and stalling reduces up to 24%. Also it reported the speedup of 7.4%. Other advantages of this strategy are reduced delay of flits and significantly less energy consumption in on-chip network.*

## *KEYWORDS*

*Directory Cache Coherence, Chip Multiprocessors, Priority Cache Coherence*

## 1. INTRODUCTION

As silicon resources becomes increasingly abundant, processor designers are able to place more and more cores on a chip with massive multicore chips on the horizon. Today's state-of-the-art general purpose chips integrate up to one hundred cores[32], while GPUs and other specialized processors may contain hundreds of execution units [24]. Commercial designs with moderate number of cores have been announced [30, 12] with shared memory architecture maintained with snoopy cache coherence protocols. Also in future generation, share memory architecture will also be the main tendency. But as the number of cores scales beyond tens, more scalable directory-based coherence protocols and on-chip interconnect (NoC) will be used.

Many coherence protocols use a subset of the classic five states MOESI model first introduced by [31]. These MOESI refer to the states of blocks in cache. Some messages need to be transferred when a cache line changes its state to another one, so transient states are needed during the transition. The speeds of different messages traverse over the NoC make the coherence protocol more complicated. For example, the GEMS [21] implementation of the MESI directory protocol – a direct descendant of the SUNfire coherence protocol – requires no less than 30 states. Another performance limitation of the protocols presented thus far is that the coherence controllers stall in several situations. In particular, the cache controllers may stall when they receive forwarded requests for blocks in certain transient states. Two examples shown in Figure 1 explain how those situations occur: in a regular implementation, for the L1cache in P1, data message (message 3) and forward message (message 4) can come in a traverse order (message 4 comes first). In this case, the cache controller may stall this message (as shown in figure 1 (a)) or transit the state to another new one (as shown in figure 1(b)). Both of those two actions will degrade the performance of whole system.

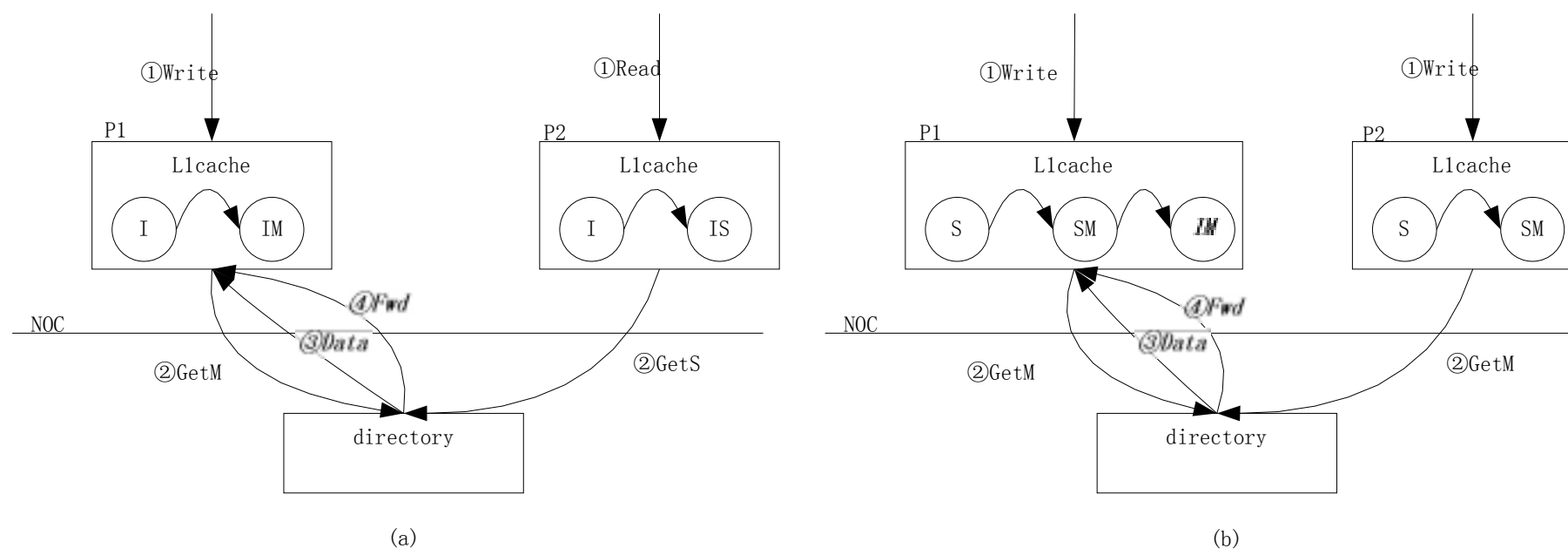

Figure 1. Case study in cache coherence protocol

On the other side, the performance of coherence protocol is dependent on the message transfer latency, which is highly dependent on the design of the on-chip interconnect (NoC) and the organization of the memory caches. For the organization of memory caches, the last level cache (LLC) or memory is typically shared across the cores and incurs a latency of access. This latency is decided by the NoC and cache coherence protocol. It is a function of distance on the chip. A recent study showed that up to 77% of the overall delay in a SoC chip can come from the interconnect in the 65nm regime [28]. Another research from [18] shows that many server applications lose almost half of their potential performance (assuming all data could be accessed at the latency of the local cache) due to the increased latency of on-chip cache access. So as the number of cores increases, NoC becomes more and more important to the whole system.

Several coherence protocol optimizations [15, 17] to address the challenge of the non-uniform latency while retaining the benefit of a shared cache have been reported in the literature [3, 6, 19]. In this paper, we propose a new cache coherence design which is called PPB (Phase-Priority Based) cache coherence. It introduces the idea of phase, provides phase-priority based coherence message and different transfer speed for those messages in NoC. PPB cache coherence achieves its goals by (1) differentiating coherence messages and giving priority to them based on the phase; (2) adopting hardware based NoC priority mechanism for efficient distributed directory-base cache coherence access in both static and dynamic NUCA system. The main power of PPB cache coherence is its simplicity and low hardware cost. It enables to facilitate the optimization of the resources in traditional NoC routers and reduce the number of stalling and unnecessary transient states in the cache coherence (as shown in figure 1).

The rest of the paper is structured as follows. In section 2, we give a short background of this work. Then in section 3, we analyze cache coherency message and translate them into a set of phases. Then we introduce PPB cache coherence protocol. Section 4 offers our evaluation of PPB cache coherence protocol. Section 5 presents the related work, and Section 6 concludes the paper.

## 2. BACKGROUND

Now CMPs are typically designed as NUCA [20] architecture, where the cache is divided into multiple banks, and access to different banks result in different latencies. In NUCA architecture, performance depends on the average latency. Figure 2 (a) presents a high-level block diagram view of a tiled CMP architecture which is regarded as the base of our design. Each process unit in the CMP contains a processor core, private L1 I&D cache, each memory unit can be either a slice of the globally-shared L2 cache (the last level cache), or a memory block. All these units are interconnected by a mesh network.

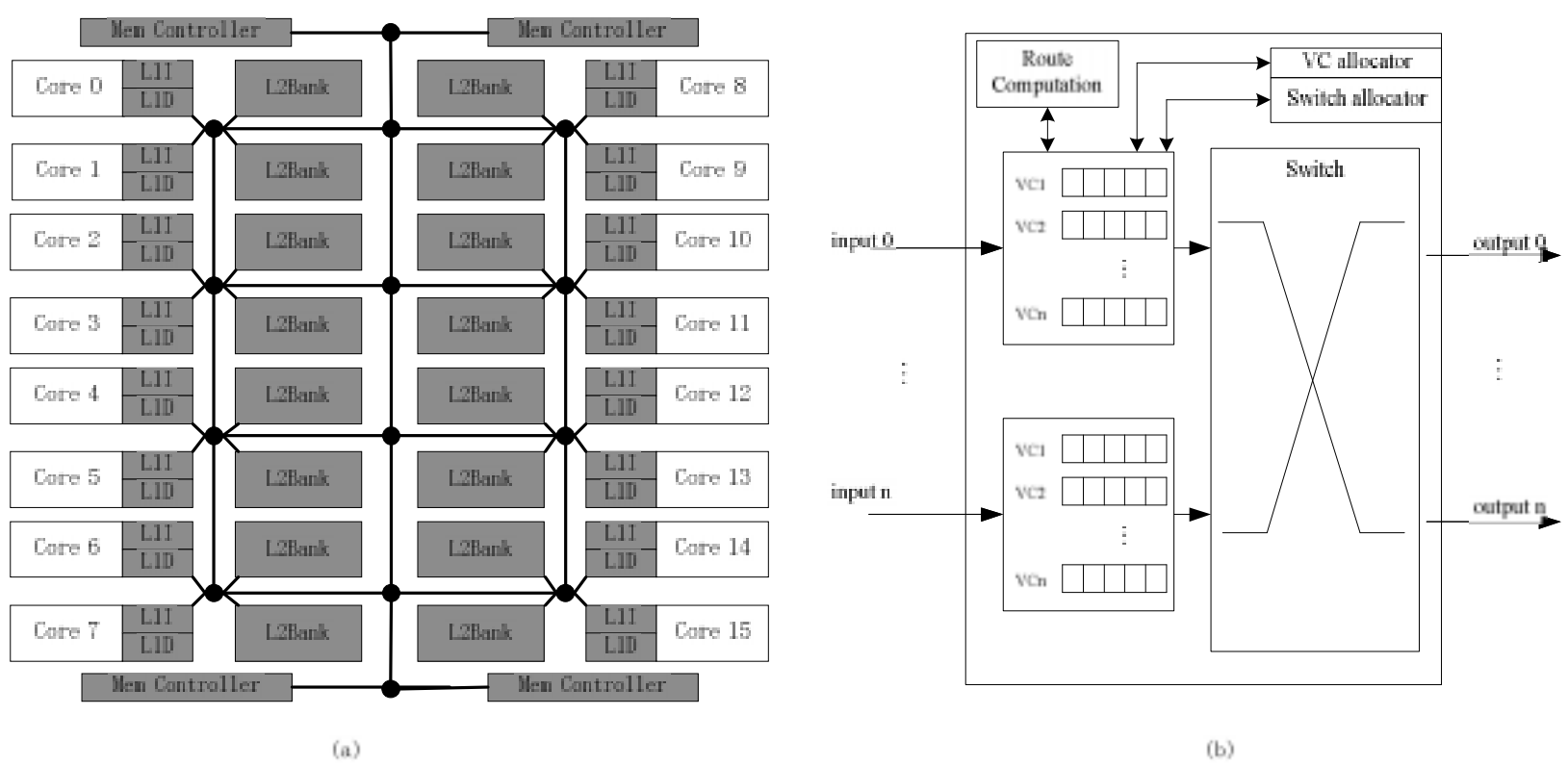

Figure 2. (a) Block diagram of the underlying CMP architecture depicting a processor with 16 cores. Coherence directory distributed to L2-bank; (b) Router microarchitecture

### 2.1. Cache coherence protocols

Cache coherence is needed to maintain the illusion of a single shared memory on a system with multiple private caches. A coherence protocol arbitrates communication between the private caches and the next level in the memory hierarchy, typically a shared cache (e.g., the L2 cache in figure 2(a)) or main memory. In this work we focus on directory-based, write-invalidate protocols. These protocols use a coherence directory to track which caches share a line, enforce write serialization by invalidating or downgrading access permissions for sharers, and act as an ordering point for concurrent requests to the same address.

In this work, the distributed directory is implemented by extending each L2 cache line with the directory information, which tracks the states of this cache line. The directory information includes a status vector, which contains the identity of cores that store this cache line in their private caches. It also contains a modified bit to indicate that this cache line is changed by one of the L1 caches. In this paper we focus on the MESI protocol which was used in SGI Orign2000 [5]. L2 cache and directory deal with incoming messages in-order for maintaining memory consistency [29].

### 2.2. Typical virtual Channel Router

Figure 2(b) illustrates a typical virtual-channel router architecture [10]. As shown in this figure, the data-path of the router consists of buffers (or virtual channels [8]) and a switch. The input buffers store flits while they are waiting to be forwarded to the next hop. When a flit is ready to move, the switch connects an input buffer to an appropriate output channel. To control the data-path, the router also contains three major control modules: a router computation unit, a virtual channel allocator, and a switch allocator. These modules determine the next hop, the next virtual, and when a switch is available for each packet/flit.

The routing operation takes four steps, namely routing computation (RC), virtual-channel allocation (VA), switch allocation (SA), and switch traversal (ST), which often represent one to four pipeline stages in modern virtual-channel routers. When a head flit (the first flit of a packet) arrives at an input channel, the router stores the flit in the buffer for the allocated virtual channel and determines the next hop for the packet (RC stage). Given the next hop, the router then allocates a virtual channel in the next hop (VA stage). After that, the flit competes for a switch (SA stage) if the next hop can accept the flit, and move to the output port (ST stage). Finally, the flit can be transferred in the link (also called LT stage).

## 3. PPB CACHE COHERENCE

In this section we firstly introduce a straightforward architecture of NUCA CMP over NoC. We briefly describe the basic CMP communication infrastructure and details of directory-based distributed coherency protocol over NoC. Then we show how to classify the coherence message, also we introduce the idea of “phase message” here. Finally we investigate how by using a priority-based NoC which is equipped with a simple priority mechanism we can drastically decrease cache access latency and improve performance of our applications. Also we present the detailed implementation of the whole system, which refers to the coherence controller and the NoC.

### 3.1. Analysis of the cache-access over NoC

Figure 3 depicts the basic cache-access transaction over NoC in our base architecture. When one core issues a read operation and encounters a L1 miss (data or instruction miss), it performs a L2 cache transaction, and this transaction is translated into multiple messages over the NoC. As shown in figure 3 (a), the L1 cache controller which is connected with C0 first issues a GetS request and sends over the NoC towards a L2 node according to the address of the block for SNUCA or after a search procedure for DNUCA. If the cache line exists in the L2 cache and it is clean, L2 cache responds this request with a Data message and sends it to the requestor through the NoC. But if the cache line is in modified states (as shown in figure 3 (b)), directory controller forwards this request to the owner which is recorded in the directory. When the owner receives this request, it forwards the data to the requestor and writes back the data to the under-level cache (here is L2 cache).

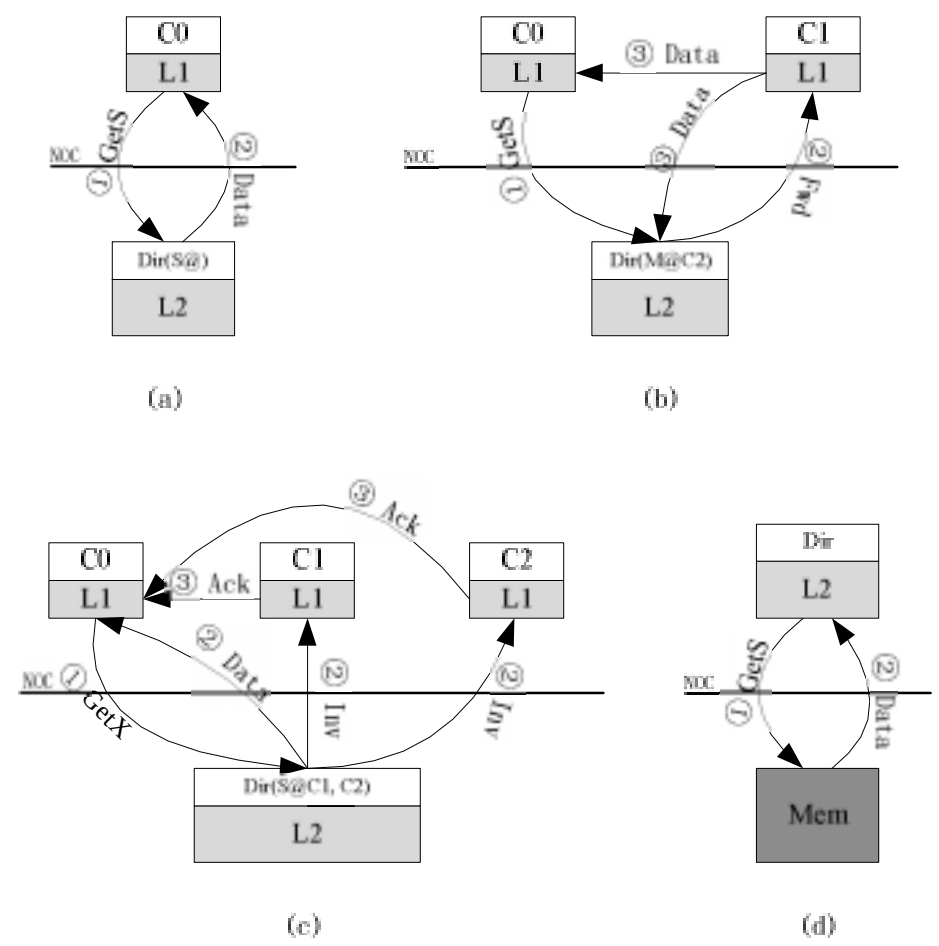

Figure 3. Cache access in CMP over NoC - (a) read transaction, (b) read transaction with the cache line in modified state, (c) write transaction, (d) memory transaction

When one core issues a write operation and encounters a L1 miss, if the cache line that is requested exists in the L2 cache and in the exclusive state (this cache line does not exist in any L1 cache), L2 cache responds this request with this cache line. This process seems similar with figure 3 (a). But if the cache line is shared by other L1 caches, as shown in figure 3 (c), the directory first sends invalidation (INV) request to the sharers (C1 and C2 in figure 3 (c)). Also it sends data to the requestor (C0 in figure 3 (c)). When the sharers receives INV message, they send acknowledge message to the requester.

In the above process, if the cache line that is requested does not exist in L2 cache, a memory transaction will be performed by the L2 cache controller. As shown in figure 3 (d), first L2 cache controller sends a GetS request to a memory node over the NoC. When memory receives this request, it responds the requestor with the data.

These actions that are depicted in figure 3 are just a basic description. Some actions are ignored in this figure, e.g. in figure 3 (b) & (c), when the requestor receives all message finally, it sends a acknowledge message to the directory controller for updating. Also we do not show the order constraint of memory consistency model. A detailed introduction can be found in [29].

## 3.2. Implementation of PPB cache coherence

Directory coherence protocol contains two important features: (1) directory is the key innovation of coherence protocol. It maintains a global view of the coherence states of each block (or cache line). In a typically directory protocol, coherence transactions typically involve either two steps (a unicast request, followed by a unicast response, as show in figure 3 (a)) or three steps (a unicast request, K>1 forwarded request, and K responses, where K is the number of sharers, as shown in figure 3 (c)). Some protocols even have a fourth step (memory transaction, responses indirect through the directory or the requestor notifies the directory on transaction completion). (2) In our directory protocol, a coherence transaction is ordered at the directory. Multiple coherence controllers may send coherence requests to the directory at the same time, and the transaction order is determined by the order in which requests are serialized at the directory.

Based on these two important features and the analysis of section 3.1, we introduce the idea of "phase" message here. On the other side, for a given block (or cache line), there are maybe more than one requests simultaneously. To get more insight into these cases, we divide the idea of "phase" into "outer phase" and "inner phase".

### 3.2.1. Outer Phase Message

For a single coherence transaction, it maybe involves either two steps or three steps, and even four steps. During this proceeding, directory is the key component to order this transaction. So we propose to differentiate these messages into before and after ordered. For the message that is first sent from L1 cache and does not arrive the directory (in other words, this message has not been ordered), we consider it as the "first phase" message. For the message that is sent by directory (here the transaction has been ordered), we consider it as the "second phase" message.

For those messages in the "second phase", we can further divide them into two categories. We consider those messages that are transferred between memory and L2 cache as the "third phase" messages. We do this not only because of the long latency of memory access, but also because memory-access is in the critical path of the whole coherence transaction (all the following requests to this cache line must be stalled until this memory transaction ends).

### 3.2.2. Inner Phase Message

Multiple local cache controllers may send coherence requests for the same cache block to the directory at the same time, and the transaction order is determined by the order in which the requests are serialized at the directory. Even this, because the latency of NoC cannot be determined, these messages which come from different transactions can also be transferred in an unregularly order. Figure 1 shows two examples. To alleviate these cases, and also to reduce the number of stalling and unnecessary transient states, we further introduce idea of "inner phase".

The "inner phase" is used to identifier these transactions that refer to the same cache block. We timestamp each coherence transaction with a phase number to indicate its order in the whole proceeding. This order is determined by the directory controller. We say that coherence transaction X is logically earlier than coherence transaction Y if X has the smaller phase number than Y. So we consider coherence transaction X as the earlier transaction and consider Y as the later one. The timestamp of each transaction is considered as its "inner phase" number.

The "inner phase" associates with the "second phase" of outer. That's because these messages in the "first phase" of outer do not be ordered, so they can be transferred in any order. Also those messages in the "third phase" of outer do not contain more than one message that refers to the same cache block at the same time in our base architecture. Besides, almost all stalling and unnecessary transient states are introduced in the "second phase".

### 3.2.3. Implementation

Figure 4 shows the identifier of phase in PPB cache coherence protocol. For each request/response message in coherence transaction, we add extra 8-bits into it to identify its phase number. The highest two bits are used to identify the outer phase number, and the rest bits are used to identify the inner phase number.

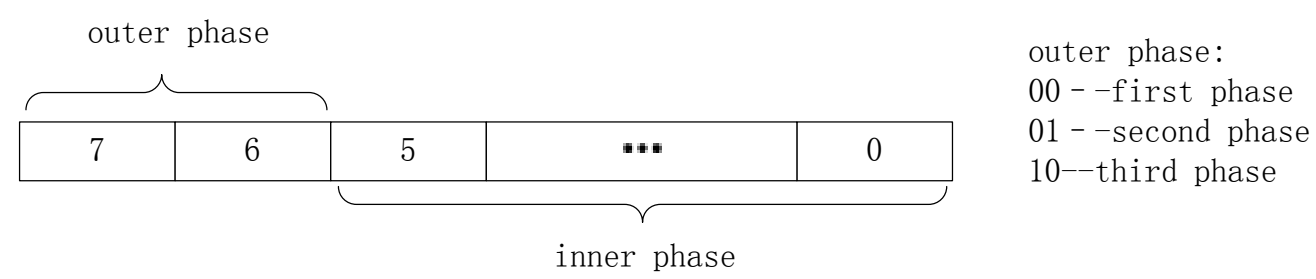

Figure 4. Phase identifier in PPB cache coherence protocol

When a coherence message is transferred in the NoC, this phase number is considered as the priority. In our mechanism, we give the following rules:

(1) Outer phase number has higher priority than inner phase number. It means if two messages contain different outer phase number, inner phase number will not be considered in our priority mechanism.
(2) For the outer phase number, the biggest number makes for the highest priority. It means that the messages in the first phase of outer have the lowest priority.
(3) For the inner phase number, when the later message arrives, its inner number increases by 1. The priority of earlier message is higher than the later message.

The above rules identify our priority arbitrator mechanism in coherence protocol. For the outer phase, we do this to make sure that the coherence request that has been ordered can finish quickly. When the messages in the first stage of outer arrives in directory, they can be done immediately if there are no earlier request to the same cache line, else they would be stalled. Moreover the messages in the second phase of outer are in-flying transactions for the directory controller, they will not be stalled again in directory theoretically. So compared with the messages in the second stage of outer, the messages in the first stage have the lower priority. We give the highest priority to the messages in the third phase of outer because the memory access is in the critical path of the whole transaction. For the inner phase, we do this to alleviate those cases like figure 1. With this inner phase mechanism, we can reduce the number of stalling and unnecessary transient states that occurs in local cache controller.

As shown in section 3.2.2, the inner phase is used between coherence transactions that refer to the same cache block. So for a given block, in order to calculate the inner phase number of later message, we need to record the inner phase number of earlier one. In our implementation, we add an extra buffer for every L2 cache bank. A fixed number of entries are involved in this buffer. Each entry in this buffer can be used to record one address of a cache line and its latest inner phase number. We use this buffer to record the messages of cache lines that were accessed recently. Because of the limited number of entries in this buffer, only the latest cache lines that were accessed can be recorded. But considering the window size of inner phase (only six bits,
which means 64 is the biggest inner phase number), it does not need too many entries in this buffer.

## 3.3. Details of router microarchitecture in NoC

Router is the key component to arbitrate those messages that are transmitted in NoC. Our base model adopts classic four-stage pipeline router microarchitecture [10], as shown in figure 2(b). Besides, in order to support our priority mechanism, a set of arbitrate mechanisms based on priority are appended in this microarchitecture. These arbitrate mechanisms are implemented in the same manner. For the incoming and waiting message, the arbitrate mechanism first traverses all of messages and then chooses the best one with the highest priority. Also a threshold value is added in this mechanism to avoid starvation of some messages that with low priority.

The first mechanism which needs arbitrator in data-path is NI (Network Interface). It is used to send message from those tiles (L1 cache, L2 cache or memory controller) to the NoC. Because of the limited resource in hardware, there are maybe multiple messages in NI in the same time. And VC (virtual channel) [8] makes this case more complexly. We need to arbitrate all messages that reside in all VCs. The next mechanism which needs arbitrator in data-path is VA (Virtual channel Allocator), here the message needs arbitrate for a VC corresponding to its output port. In this stage, we need to choose the message with the highest priority among all messages in all VCs and allocate output VC for it firstly. Then upon successful allocation of output VC, the message advances to the switch allocation (SA) stage where it arbitrates for the switch ports. Here we need to traverse all messages that come from all inputs and all VCs, and then we can choose the best candidate that with the highest priority. During the above process of arbitration, priority is the first consideration. But if there are more than one message which contain the highest priority, round-robin mechanism is used.

## 4. EVALUATION

We evaluate the scheme proposed in section 3 through simulation. In subsection 4.1 we present the simulation environment. Then in the following subsections, we present the simulation results that demonstrate the advantage of our mechanism.

### 4.1. Methods and Workloads

Base System. We evaluate our mechanism through full simulation of a 16-way chip multiprocessor (CMP) with private L1 caches and a shared L2 cache (as shown in figure 2(a)). The system contains a 14-node mesh interconnect that use 16-byte links with X-Y routing; four memory controllers are used to access main memory. For the buffer that is used for inner phase, it contains 32 entries. Detailed system parameters are shown in Table 1.

Simulation Methods. Our mechanism has been simulated using the Simics [23] simulation infrastructure from Virtutech and GEMS[21, 33] toolset from Wisconsin’s Multifacet group. We modified the performance models of GEMS, but left Simics full-system infrastructure unchanged. All of these benchmarks run on unmodified Solaris 10. The GARNET [2] network model was used to capture the detailed aspects of the interconnection network. GARNET is a cycle-accurate interconnect model that models a detail packet-switched [9] router pipeline including VCs, buffers, switches and allocators. GEMS, alone with GARNET, provide a detailed memory system timing model. The Orion[22] was used as our power model.

Table 1. Base system parameters

| | |
|---|---|
| Core | 1-issue in-order SPARC core, 16 |
| L1Cache | Private, 32K I-cache, 32K D-cache,4-way set associative, 64byte blocks, 2-cycle latency |
| L2Cache | 8M shared cache, 8-way set associative, 12-cycle latency |
| Network | 3X3 mesh, flit width is 128-bit, 4 cycle router pipeline, 5 virtual channels, 2-cycle link latency, Dimension-ordered X-Y routing algorithm |

Workloads. SPLASH-2[34] benchmark suite is used for experiments. SPLASH-2 is a suite of scientific evaluations for the past two decades. The number of threads spawned in a multithread workload is the same number of hardware threads so each thread is statically mapped to a hardware thread. We use all SPLASH-2 applications, and the datasets used for SPLASH-2 applications are summarized in Table 2. For each application, the same dataset is used throughout all process generations. To address the variability in parallel workloads, we simulate each design point multiple times with small, pseudo-random perturbations of request latencies to cause alternative paths to be taken in each run [1]. We average the results of all runs.

Table 2. SPLASH-2 datasets in evaluation

| **Application** | **Dataset** | **Application** | **Dataset** |
|---|---|---|---|
| barnes | 16K particles | water | 4K molecules |
| fft | 1024K points | cholesky | Tk17.O |
| fmm | 16K particles | radiosity | room |
| Lu | 512*512 matrix | raytrace | car |
| ocean | 258*258 grids | volrend | head |
| radix | 8M intefers | | |

## 4.2. Performance Analysis

In this subsection we report our simulation results and analyze the effectiveness of our mechanism. All results are with respect to the original executable running with standard MESI cache coherence protocol that was used in SGI Orign2000 [5].

### 4.2.1. Effect on transient states and stalling

Firstly we measure the number of unnecessary transient states and stalling in coherence transaction. Both of them are the key actions to affect the performance of application, which has been shown in section 1 and figure 1. Figure 5 shows the average number of transient states and stalling reduction in our scheme. For the transient states, we just choose one case that occurs in L1 cache: we just count the number of INV messages that are received by cache lines that in IS state. For the stalling, we just statistic the number of stalling that occurs in L2 cache.

NoC is the major factor to generate these unnecessary transient states and stalling. Adding the phase number, especially the "inner phase" number into these messages and adding priority support in NoC, our mechanism can reduce the number of these unnecessary cases efficiently. As shown in figure 5, we can see that PPB mechanism can reduce transient states by 24% on average. Also as expected, the number of stalling in L2 cache can achieve a reduction of 19% on average. The maximal reduction reached 57% in Cholesky for transient states, and 32% in barnes for stalling. All these results identify the effect of our mechanism.

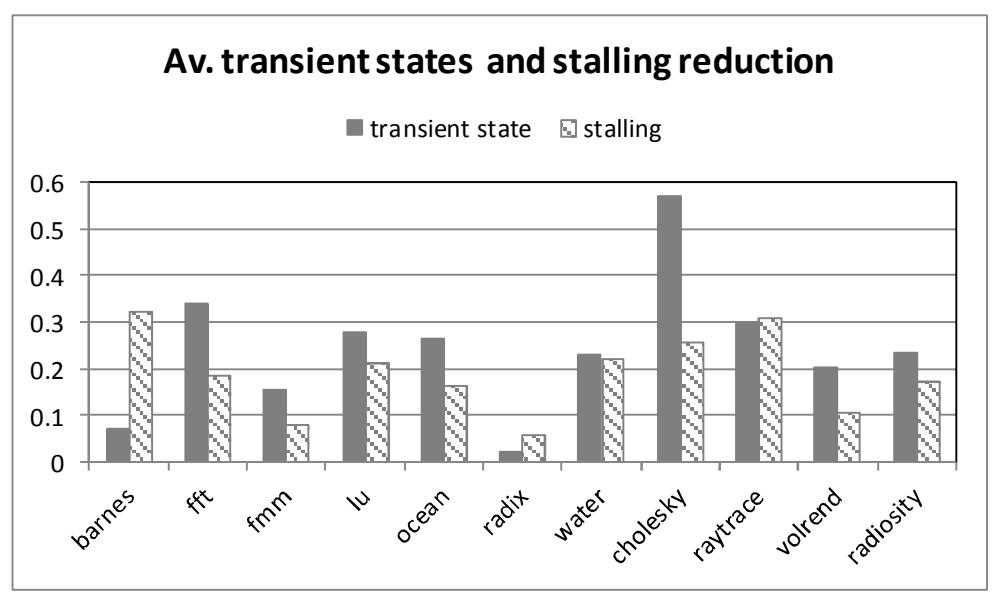

Figure 5. Average number of transient states and stalling reduction in coherence transactions

### 4.2.2. Effect on delay

Another objective of our evaluation focuses on the delay of read/write transactions. The delay of a transaction is calculated from the time that the core sends the read/write request until it receives the response. In other words, the overall delay of a transaction consists of the dealing time at local cache controller, the queuing time at the L1/L2/memory interface, the NoC delay of all traversed messages that are introduced by this request.

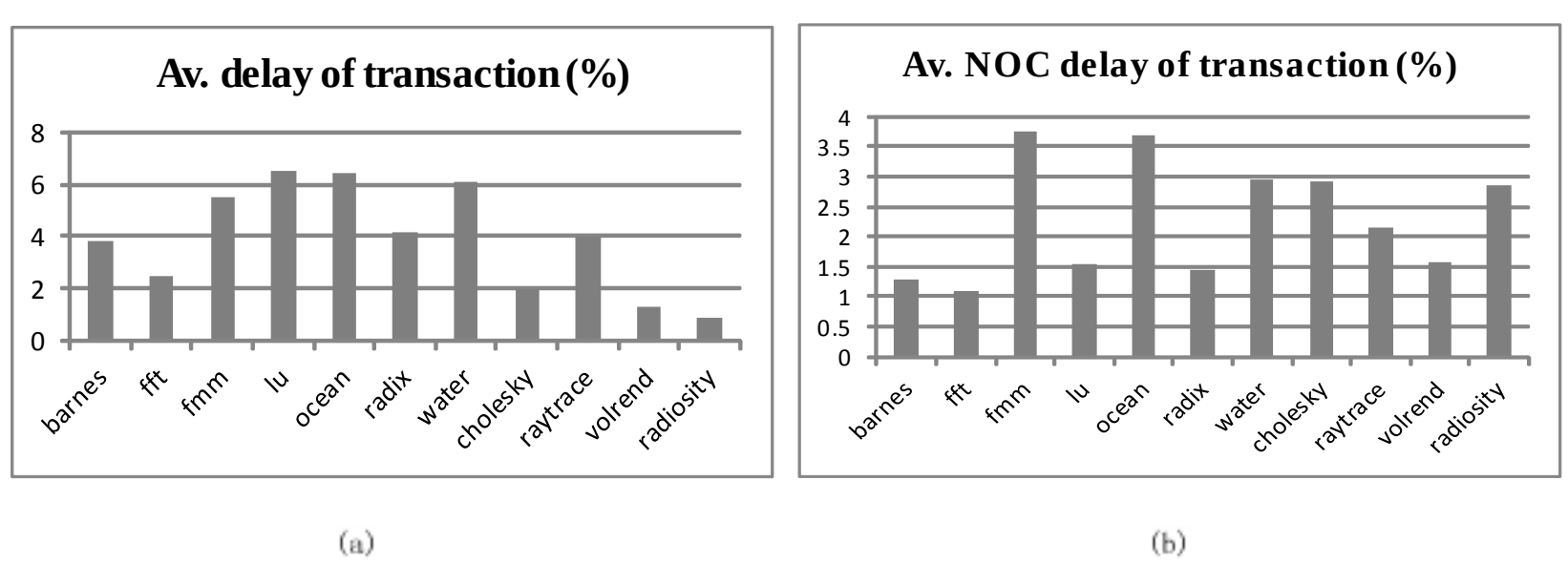

Figure 6. (a) Average delay reduction of a transaction. (b) Average NoC delay reduction of a transaction

Figure 6(a) quantifies the average delay reduction that can be achieved by our mechanism. From this graph, PPB mechanism can achieve a delay reduction of 4% on average. And the maximal delay reduction reaches 6.4% in LU. To evaluate the NoC in advance, we also measure the average NoC delay reduction (as shown in figure 6(b)). From this figure we can see that PPB mechanism reduces NoC delay, on average, by 2.3% for our workloads. Besides, comparing the results of figure 6(a) with 6(b), we can get the average queuing delay reduction of coherence transaction at the L1/L2/memory interface. So we can report that our mechanism can improve the performance from all related aspects of message traverse.

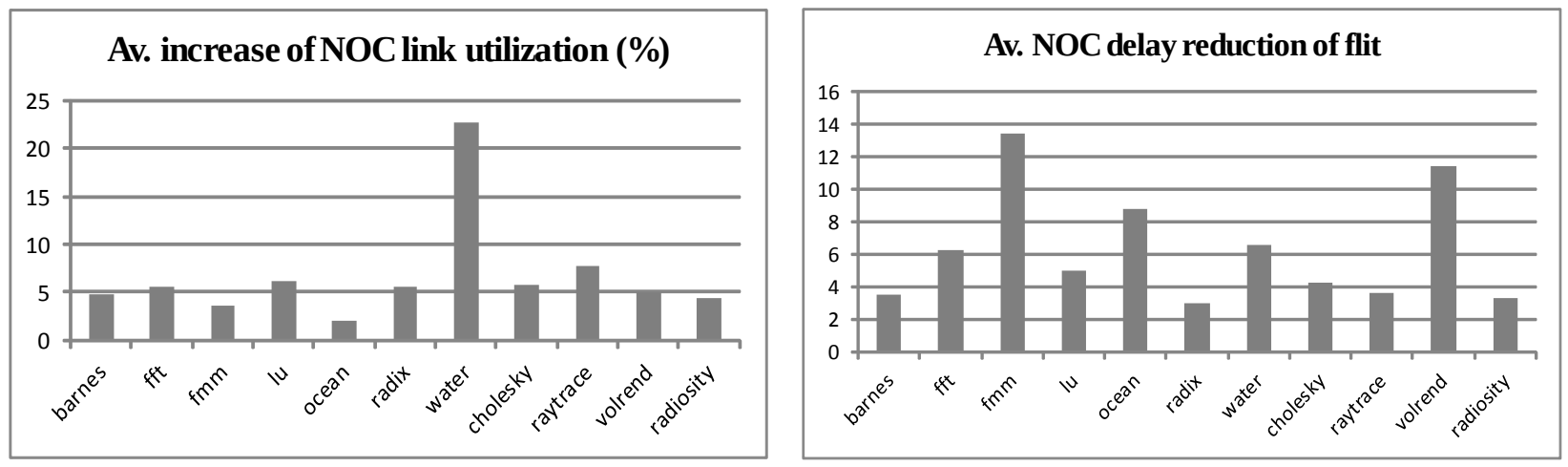

(a) (b)

Figure 7. (a) Average link utilization improvement in NoC. (b) Average latency reduction of flit in NoC

To evaluate the effect of NoC in a fine granularity, we also measure the NoC link utilization, which in terms of the number of flits that are transferred per cycle in NoC. The result is shown in figure 7(a). From this graph we can find that our mechanism can improve the link utilization by 6.6% on average. Even highest improvement achieves 22.7% in Water. Figure 7(b) depicts the latency reduction of flits that are traversed in NoC. For 16 tiles, PPB mechanism exhibits an average reduction of 6.3% for our workloads.

### 4.2.3. PPB cache coherence Performance Improvement

In this subsection we examine the total application speedup that achieves by our mechanism. The results are depicted in figure 8. We observe that PPB mechanism improves the overall system in all benchmarks. The best improvement is achieved in Radiosity, which boosts overall system performance by 7.4%. The system performance improvement strongly depends on the amount of coherence transactions, the amount of coherence transaction delay reduction and the number of unnecessary transient states and stalling in coherence controller.

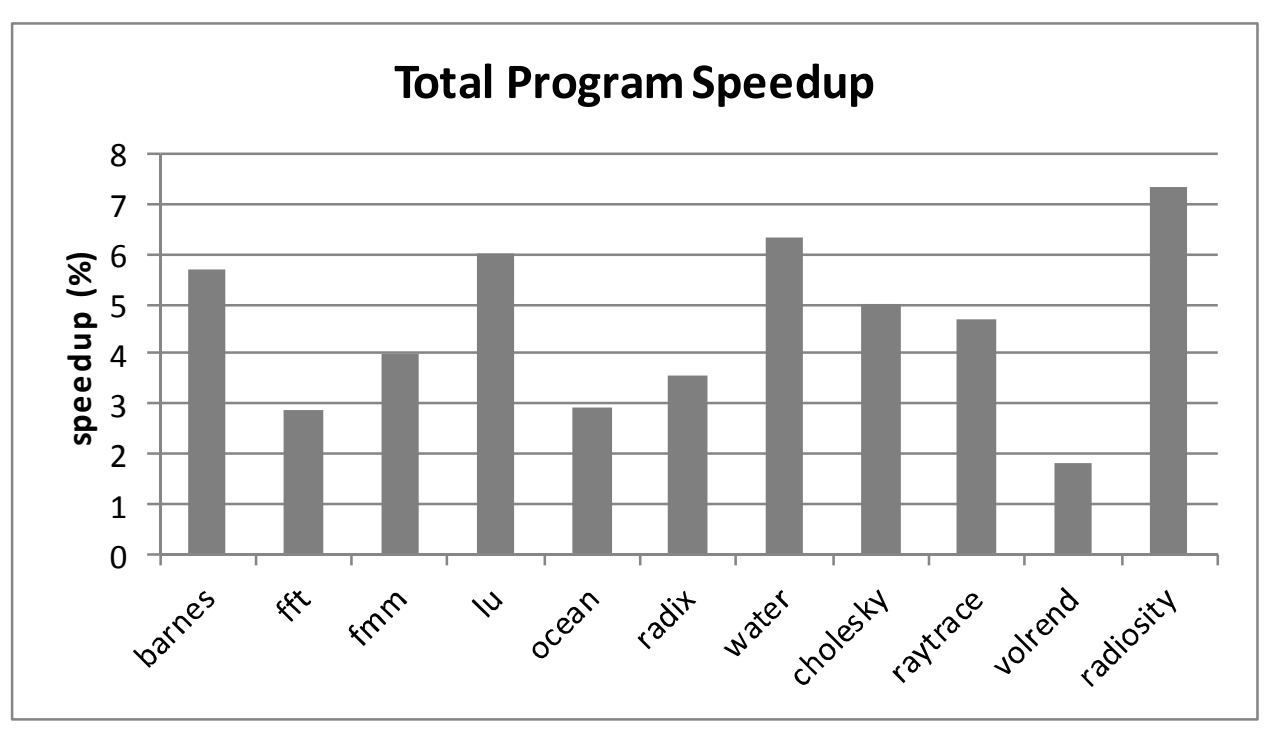

Figure 8. Overage program speedup by using PPB mechanism

### 4.2.4. Effect on dynamic energy

Another consideration of our scheme is energy consumption. Since the introduction of priority, some arbitrators in router must be changed to adaptive this case, which will introduce extra energy consumption for our base system. But network energy consumption can be saved by reducing the number of coherence transactions generation and the delay of data communication.

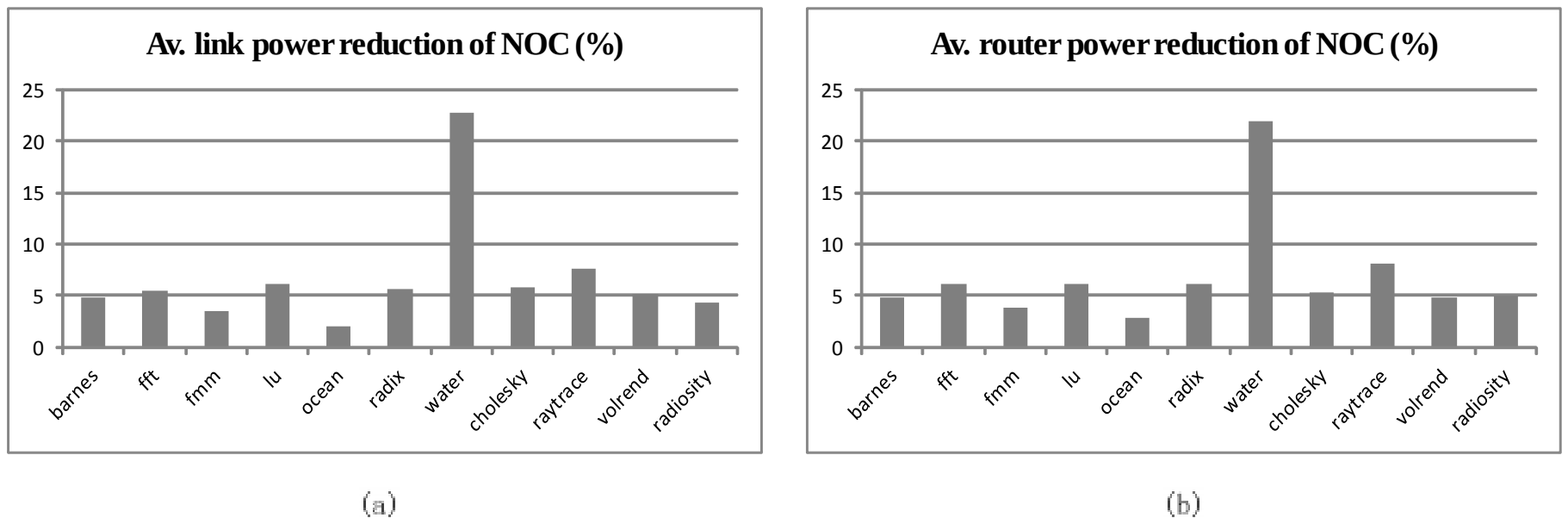

Figure 9. Average dynamic power consumption reduction of (a) link, (b) router power

Figure 9 reports dynamic energy consumption normalized to standard MESI protocol for SPLASH-2 benchmark. According to the graph 9(a), the reduction of dynamic link energy consumed by PPB mechanism can achieve the highest 22.7% in Water benchmark, and 6.6% on average. For router, as shown in figure 9(b), it consumes 6.7% less energy over standard MESI protocol. All this reduction comes from the reduction of the number of coherence messages. As the number of unnecessary transient states and stalling degrade, some messages that are introduced by these cases eliminate, which reduce the energy consumption indirectly.

## 5. RELATED WORK AND DISCUSSION

In this section we briefly describe the most closely related previous work. There has been substantial prior work in the area of cache coherence optimization in the context of CMP [3, 4, 6, 7, 13, 17, 25, 26]. The majority focused on in-protocol optimizations, releasing consistency model and reducing complexity of implementation. Bolotin et al. [4] propose a similar mechanism, but it just prioritizes control packets over data packets in the NoC. The main idea in [4] is to make sure that the short message can be transferred faster. But the priority in PPB is used to make sure that the transaction that has been ordered can be done quickly. This is the main difference between [4] and our work. Also [4] does not show the detail implementation of this mechanism in NoC. Eisley et al. [11] addressed the cache coherency problem in CMP and proposed to alter the standard directory-based system by directories implemented inside NoC router. This approach enables to reduce memory access delay but requires additional storage and a more sophisticated router architecture to perform directory-related manipulations on every packet at every hop. Besides, it can make overall latency variable due to potential deadlock recovery.

In the other side, ARMCO [13] allows data to be sourced from any sharer (not necessarily the owner) via directory L1-L1 communication, with the goal of leveraging locality of access. POPS [17] decouples data and metadata and provides localized data and metadata access for both shared data and private data. Also there are some other proposals that optimize coherence actions for directory based cache coherence [16, 14]. Our mechanism can be embedded in all above proposals and cooperation with them well. A detailed performance measure for these will be done in our next work.

## 6. CONCLUSION

This paper contributes PPB cache coherence protocol, a protocol optimization based on standard MESI protocol. In this proposal, we explore the detailed action of coherence transaction and figure out the problem of the unnecessary transient states and stalling in the implementation of modern coherence protocol. Then we introduce the idea of "phase" for the messages in coherence transaction to deal with these problems. Detailed simulations show an impressive reduction in the number of unnecessary transient states (24% on average) and stalling (19% on average). Besides, our mechanism further enriches the performance of NoC. Specifically, our experimental results indicate that, when using PPB cache coherence protocol, the average delay reduction of flit can up to 6.3%. The simulations also demonstrate a substantial total system speedup in all simulated benchmarks. And the energy consumption of NoC link and router are reduced because the number of coherence transactions decreases. Finally, our scheme can be embedded in most of modern coherence protocols (such as [13, 14, 16, 17]). A detailed performance measure for these will be done in our future work.

## ACKNOWLEDGMENTS

This work is supported financially by the National Basic Research Program of China under contract 2011CB302501, the National Natural Science Foundation of China grants 60970023, the National Hi-tech Research and Development Program of China under contracts 2012AA010902 and 2012AA010901.

\

**Authors**

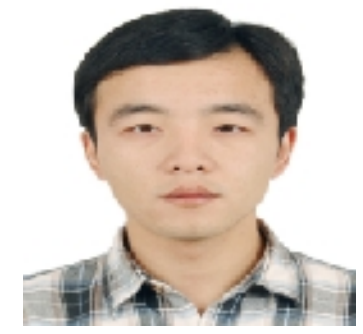

**Gongming Li** is a D.Sc. student at the Department of Computer Science and Technology in University of Science and Technology of China. His research is concerned with transactional memory, thread-level speculation, cache coherence and computer simulator.

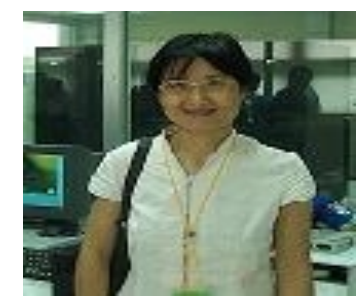

**Hong An** is a Professor of the Department of Computer Science and Technology at U niversity of Science and Technology of China. Her research is concerned on computer architecture and GPU architecture.